\documentstyle[12pt]{article}
\newcommand{\be}{\begin{equation}}
\newcommand{\ee}{\end{equation}}
\begin{document}
\title{ A simple solvable model of body motion in a one-dimensional resistive
medium}
\author{{\bf M. I.  Molina\footnote{mmolina@abello.dic.uchile.cl}}
\vspace{1 cm}
\and
\and
Facultad de Ciencias, Departamento de F\'{\i}sica, Universidad de Chile\\
Casilla 653, Las Palmeras 3425, Santiago, Chile.}
\date{}
\maketitle
\baselineskip 24 pt
\vspace{0.5in} 
\begin{center}
{\bf Abstract}
\end{center}

\noindent
We introduce and solve in closed form a simple model of a macroscopic
body propagating in a one-dimensional resistive medium at temperature $T$.
The assumption of completely inelastic collisions between the body and
the particles composing the medium leads to a resistive force that is
opposite and proportional to the square of the body's velocity.
\vspace{1cm}

\noindent
\centerline{\em Key words:\ \ air drag, collisions}\\
\centerline{\em PACS:\ 45.20.Dd , 45.50.Tn \ }

\newpage

The topic of macroscopic bodies moving through resistive media, 
such as air or viscous fluids, gives rise to one aspect that
students of Introductory Physics courses find rather mysterious:
The origin of the `force law' that describes the effective force
on the moving body as it propagates through the resistive medium.
The student is usually told that the effective force on the body is
either proportional to the speed of the body or to the square of the
body's speed, according to whether the body has, or does not have, a small
cross--sectional area,
or whether it is moving at low or high speeds$^{1}$. Under
further questioning, the instructor might tell the student that these
`laws' are based on `experimental observations' which are difficult
to obtain analytically. There are basic models, however,
that show in a simple manner how the energy and momentum exchange between the
moving body and the particles composing the resistive medium lead to some
of these `force laws'. In this article, we present an extremely simplified
model that leads to a very well--known `force law': a resistive force that is
opposite and proportional to the square of the body's velocity: $F = -\gamma\ V^{2}$.

Consider a body, represented by a heavy point `particle' of mass $M_{0}$ and
initial speed $V_{0}$, that propagates inside a one--dimensional medium 
composed of identical point particles of mass $m$, with $m \ll M_{0}$ which are in thermal
equilibrium at temperature $T$ (Fig.1). We will consider here the `short' time scale where the
body does not have enough time to reach thermal equilibrium with the surrounding medium.
The `brownian motion' case where the body is in thermal equilibrium with the medium have been
nicely discussed by de Grooth$^{3}$.
Let us denote by $v_{j}$, the velocity of the $j$th medium
particle. Since the medium is one-dimensional the particles can be labelled unambiguously. For instance,
the particles to the right of the body could be labelled by odd values of $j$, while the ones
to the left, by even $j$ values. Because of thermal equilibrium the $\{v_{j}\}$ are random quantities whose values 
are taken from a gaussian distribution of width proportional to the medium temperature $T$.
We will assume for simplicity that the body undergoes completely inelastic collisions
with the medium particles.

After the first collision we have, because of momentum conservation,
\[
M_{0} V_{0} + m v_{1} = ( M_{0} + m ) V_{1}
\]
i.e., the speed of the body after its first collision is
\[
V_{1} =  \left( {M_{0}\over{M_{0} + m}} \right) V_{0} + \left( {m\over{M_{0} + m}} \right) v_{1}, 
\]
where $v_{1}$ denotes the velocity of the medium particle with which the body collides {\em first}
(this particle could come from the left or right of $M$). Some time afterwards, the body
(now with mass $M_{0} + m$) will suffer a second inelastic collision from which will emerge with velocity:
\[
V_{2} = \left( {M_{0}\over{M_{0} + 2 m}} \right) V_{0} + \left( {m\over{M_{0} +2 m}} \right) (v_{1} + v_{2}).
\]
where $v_{2}$ is velocity of the medium particle who suffers the {\em second} collision with $M_{0}$, 
and so on. After $n$ of these collisions, the speed of the body will be
\[
V_{n} = \left( {M_{0}\over{M_{0} + n m}} \right) V_{0} + \left( {m\over{M_{0} + n m}} \right) \sum_{j=1}^{n} v_{j},
\] 
where we remind the reader that the $\{v_{j}\}$ are random with $\langle v_{j} \rangle = 0$,
$\langle v_{j}^{2} \rangle = k T/m$ and $\langle ... \rangle$ denotes a thermal average. This
implies,
\be
\langle V_{n} \rangle = \left( {M_{0}\over{M_{0} + n m}} \right) V_{0}. \label{eq:vn}
\ee
On the other hand,
\begin{eqnarray}
\langle V_{n}^{2} \rangle & = & {M_{0}^{2} V_{0}^{2}\over{(M_{0} + n m)^{2}}} +
			{2 m M_{0} V_{0} \over{(M_{0} + n m)^{2}}} \left\langle \sum_{i} v_{i} \right\rangle +
		{m^{2}\over{(M_{0} + n m)^{2}}} \left\langle \sum_{i,j} v_{i} v_{j} \right\rangle\nonumber\\
		          & = & \langle V_{n} \rangle^{2} + {n m k T\over{( M_{0} + n m )^2}}\nonumber \\
			  & = & \left( 1 - {k T\over{M_{0} V_{0}^{2}}} \right)\langle V_{n} \rangle^{2}  +
			{k T\over{M_{0} V_{0}}} \langle V_{n} \rangle. \label{eq:v2n}	
\end{eqnarray}
We note that, as the number of collisions tends to infinity (i.e., after a `long' time),
$\langle V_{n}^{2} \rangle \rightarrow (k T /M_{0} V_{0}) \langle V_{n} \rangle =
(k T/M(n))$, where $M(n) = M_{0} + n m$ is the effective
body mass after $n$ collisions. This is nothing else but equipartition: $M(n) \langle V_{n}^{2} \rangle \rightarrow k T = m \langle v^{2} \rangle$,
where $v$ is the velocity of a medium particle.

If we now assume that $\rho$, the density of medium particles per unit length 
is constant, then we can express $n$ as $n = \rho\ x$ where $x$ is the
distance travelled by the body between its first and {\em n-th} collision.
We are assuming here, as in hydrodynamics, that an element of length
$\Delta x$ while `small' will contain a great number of medium particles.
By re-expressing $n$ in terms of $x$ in (\ref{eq:vn}), we can express
the average velocity of the body after it has travelled a distance $x$ as
\be
\langle V (x) \rangle = \left( {M_{0}\over{M_{0} + \rho\ m\ x}} \right) V_{0},\label{eq:vx}
\ee
and the average of the velocity squared as
\be
\langle V(x)^{2}\ \rangle = \left( 1 - {k T\over{M_{0} V_{0}^{2}}} \right)\langle V(x) \rangle^{2}  +
		        	{k T\over{M_{0} V_{0}}} \langle V(x) \rangle. \label{eq:v2x}
\ee
The average velocity decreases monotonically with distance. Its explicit time dependence can be
found from (\ref{eq:vx}): $dx/dt = M_{0} V_{0}/( M_{0} + \rho m x )$, which can be integrated to give
\be
{X(t)\over{X_{0}}} = -1 + \sqrt{ 1 + 2 (t/t_{0})} \label{eq:x}
\ee
where $X_{0} \equiv M_{0}/(\rho m)$ and $t_{0} \equiv X_{0}/V_{0}$ constitute natural length and
time scales. Finally, after replacing (\ref{eq:x}) into (\ref{eq:vx}), or by direct differentiation
of (\ref{eq:x}), one obtains
\be
{\langle V(t) \rangle\over{V_{0}}} = {1\over{\sqrt{1 + 2 \ (t/t_{0})}}}\label{eq:vt}
\ee
and
\be
{M(t)\over{M_{0}}} \ = \ 1 + \left({\rho m\over{M_{0}}}\right) X(t) \ = \ \sqrt{1 + (2\ t/t_{0})}\label{eq:Mt}
\ee
is the effective body mass as a function of time. In Fig.2 we show $M(t)$, $X(t)$ and $V(t)$, all of which
diverge at long times.

\noindent
{\em Average resistive force}.\ \ As the body propagates, it is being hit from front
and back by medium particles which stick completely to it after colliding.
This accretion process is rather akin to the {\em opposite}
process that occurs in the propulsion of a rocket engine: instead of expelling matter our body absorbs it.
One process is the time-reversal of the other.
The average effective force on the body can be directly computed from $\langle F \rangle = M(t) d \langle V(t) \rangle/dt$.
From Eqs.(\ref{eq:vt}) and (\ref{eq:Mt})
one obtains:
\be
\langle F \rangle = - {M_{0} V_{0}\over{t_{0}}} {1\over{1 + 2 (t/t_{0})}}
\ee
which can be recast as
\be
\langle F \rangle = -\gamma \ V(t)^{2}
\ee
with $\gamma \equiv \rho m$.

Another way to compute $\langle F \rangle$ is to start from conservation
of momentum during an infinitesimal collision,
$M(x) V(x) + dM(x) v = ( M(x) + dM(x) ) (V(x) + d V(x))$.
This implies that the instantaneous force on the body is
\be
M {d V\over{d t}} = - \left({d M\over{d t}}\right) (V - v)
\ee
where $v$ is random. The average force on the moving body will then be
\be
\langle F \rangle = - \left({d M\over{d t}}\right) \langle V \rangle =
- \left({d M\over{d x}}\right) \langle V \rangle^{2}.
\ee
Since $M(x) = M_{0} + \rho m x$, we now have $\langle F \rangle = -\gamma \ \langle V \rangle^{2}$,
with $\gamma \equiv \rho m$ as before.

\noindent
{\em Stopping power}.\ \ The stopping power $\langle S \rangle$ of a medium is defined by the
average energy per unit length, lost by a projectile while traversing a resistive medium:
\be
\langle S \rangle = \left\langle {d E\over{ d x }} \right\rangle =
{d \over{ dx }}\left\{ {1\over{2}} M(x) \langle V(x)^{2} \rangle \right\}.
\ee
From Eqs.(\ref{eq:v2x}), (\ref{eq:vx}) and
the relations $(d/dx) \langle V(x) \rangle =
-(\rho m /M_{0} V_{0}) \langle V(x) \rangle^{2}$ and $M(x) = M_{0} V_{0}/\langle V(x) \rangle$,
we have
\begin{eqnarray}
\langle S \rangle & = & -{\rho m \over{2}} \left( 1 - {k T\over{M_{0} V_{0}^{2}}} \right) \langle V(x) \rangle^{2}\nonumber\\
		  & = & -{\rho m \over{2}} \left( 1 - {k T\over{M_{0} V_{0}^{2}}} \right)\left({M_{0}\over{M_{0} + \rho m x}} \right)^{2} V_{0}^{2}\nonumber\\
		  & = &  -{\rho m \over{2 M_{0} }} {( 2 E - k T )^{2}\over{( M_{0} V_{0}^{2} - k T )}}.
\end{eqnarray}
Figure 2 shows $\langle S \rangle$ as a function of distance traversed inside the
medium for several temperatures.
Note that, since we are assuming the body's initial kinetic energy to be
higher than the average thermal energy, the body will always lose energy to the medium,
on average. This energy loss becomes smaller and smaller as the body traverses the
medium. Only after an infinite amount of time, or distance travelled, will the
body's average energy loss reach zero, where a thermalization process will occur.

\noindent
In summary, we have introduced and solved in closed form the dynamics of a simple
model of a body moving through a resistive medium. We find that the effective
resistive force is opposite and proportional to the square of the body's speed$^{2}$.
\vspace{2cm}

\newpage
\centerline{{\bf References}}
\vspace{2cm}

\noindent
$^{1}$ Raymond A. Serway, {\em Physics for Scientists and Engineers with
modern physics}, 2nd. ed. (Saunders College Publishing, Philadelphia, 1986),
pp. 115--118.
\vspace{1cm}

\noindent
$^{2}$ A related model, where the collisions between the body and the
medium particles is assumed to be completely elastic, leads to the same
force law (but a different $\gamma$) and is reported in:\ M. I. Molina, ``Body Motion in a
One-Dimensional Resistive Medium'', M.I. Molina,
Am. J. of Phys. {\bf 66}, 973--975 (1998).
\vspace{1cm}

\noindent
$^{3}$ Bart G. de Grooth, ``A Simple model for Brownian motion leading to the Langevin equation'',
Am. J. Phys. {\bf 67}, pp. 1248--1252.
\newpage

\centerline{{\bf Figure Captions}}
\vspace{2cm}  

\noindent
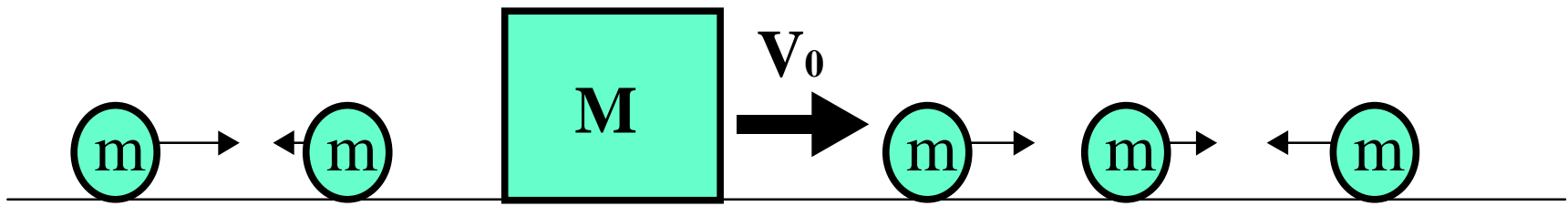
FIG 1:\ \ Macroscopic body of mass $M_{0}$ propagating inside a one-dimensional
resistive medium composed by identical particles of mass $m << M_{0}$ in thermal
equilibrium at temperature $T$.
\vspace{1cm}

\noindent
FIG 2:\ \ \ Effective body mass, average velocity and distance travelled as a
function of time, for body moving through our resistive medium
($S_{0}\equiv \rho m V_{0}^{2}/2$).
\vspace{1cm}

\noindent
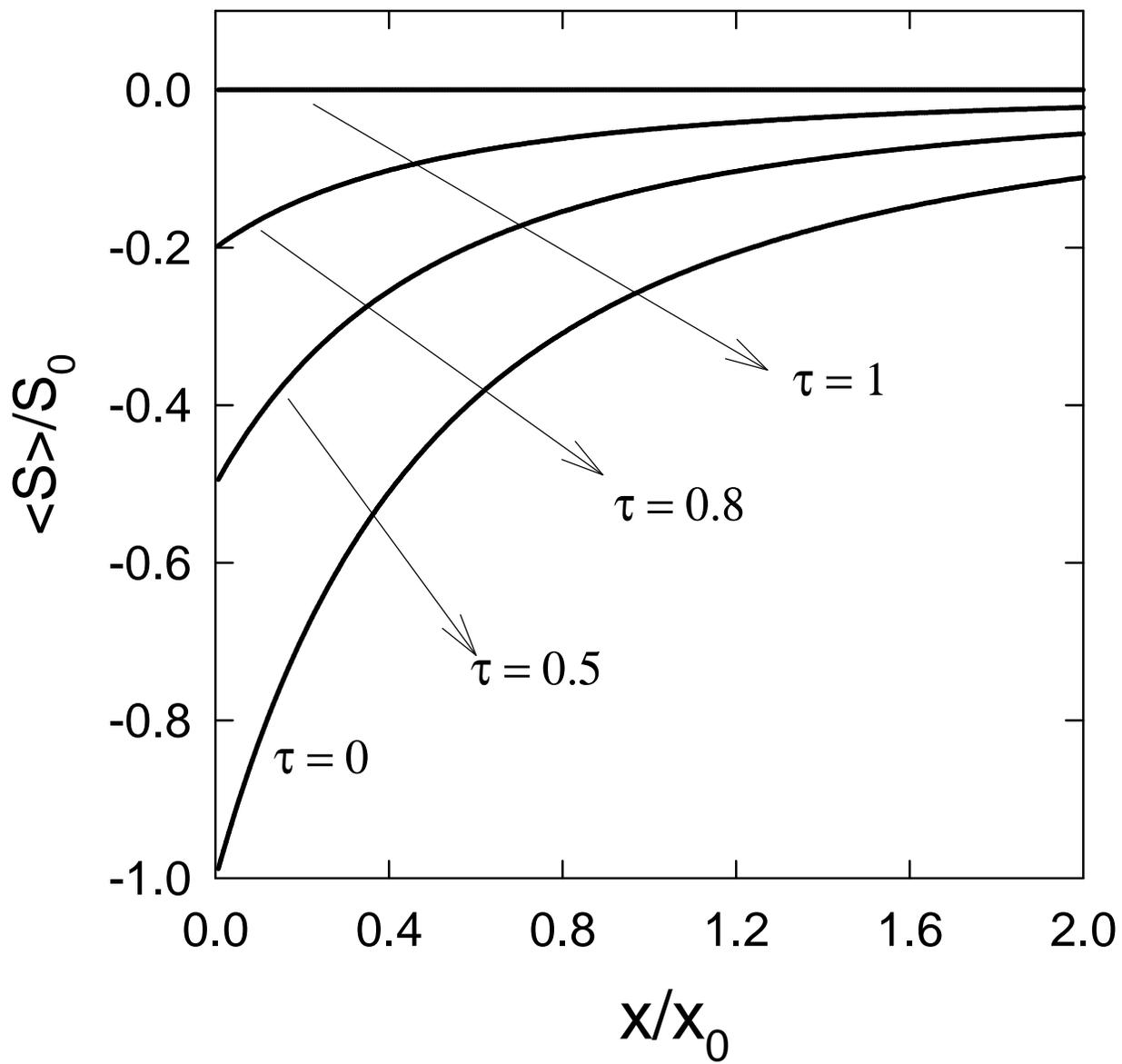
FIG 3:\ \ \ Stopping power of the one-dimensional resistive medium as a function
of the distance traversed by the body, for several medium temperatures. 
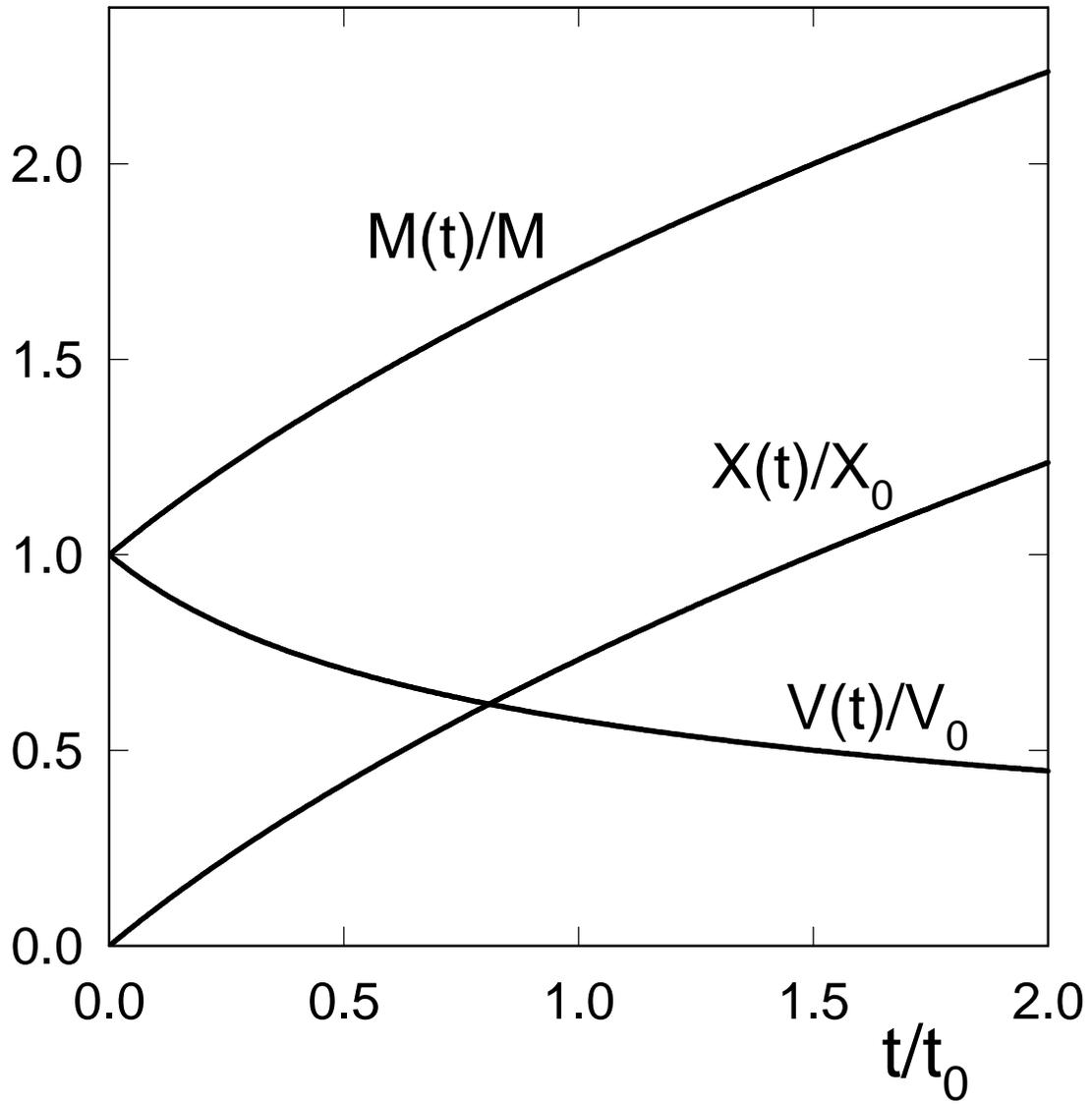
\end{document}